\documentclass[11pt,a4paper]{article}
\pdfoutput=1

\usepackage{graphicx}
\usepackage{caption}
\usepackage{subcaption}
\usepackage{bm}% bold math
\usepackage{amsmath}
\usepackage{amssymb,xfrac,comment}
\usepackage[bbgreekl]{mathbbol}
\usepackage{cite,color,float}
\usepackage{enumitem}

\def\be{\begin{equation}}
\def\ee{\end{equation}}
\def\bea{\begin{eqnarray}}
\def\eea{\end{eqnarray}}

\def\rd{\mathrm{d}}
\def\ri{\mathrm{i}}
\def\I{\mathrm{I}}
\begin{document}

\pagestyle{plain}

\begin{center}
\textbf{\large Lost in Normalization}
\vskip .5cm
Narges Vadood%~\footnote{n.vadood@alzahra.ac.ir}
~~~~~~and~~~~~~~
Amir H. Fatollahi \footnote{Corresponding Author: fath@alzahra.ac.ir}
\vskip .3cm
\textit{
Department of Physics, Faculty of Physics and Chemistry, \\ Alzahra University, Tehran  1993891167, Iran
%Department of Physics, Alzahra University, Tehran 1993891167, Iran
}
\end{center}
\begin{abstract}
\noindent
The consequences of the gauge-coupling dependent normalization-factor 
of $1/g^{\alpha}$ in the transfer-matrix of 2d U(1) lattice gauge theory 
are explored. It is seen by the $\alpha=1$ choice that the 
lowest energy develops a minimum at coupling $g_*=1.125$, 
leading to a \textit{multi-valued} Gibbs energy similar to the systems
with the first-order phase transition. 
It is argued how the $1/g$ normalization may be regarded as
a lost normalization in the commonly used change of variable to 
the dimensionless angle-variables. 
Based on the continuum limit at the next-leading order and 
the Ostrogradsky formulation of higher-order 
time-derivatives theories, it is argued that the spectrum at continuum is 
compatible only with the $\alpha=1$ choice. 
\end{abstract}

\noindent\textbf{Keywords:} Lattice gauge theories; Transfer-matrix method; 
Energy spectrum
\vskip.5cm
%\textbf{pacs numbers}: ??

%\section{Introduction}

The main concern of the present note is the neglected \textit{gauge-coupling dependent} 
normalization of the lattice gauge transfer-matrix, through which 
it is shown a first-order phase transition of the model is lost. To raise the issue, let us begin 
with the familiar case of the partition function of a free particle by means of the time-sliced Euclidean action 
(units $\hbar=1$)
\begin{align}\label{01}
Z_\mathrm{particle}^\mathrm{1d} = \int_{-L/2}^{L/2} \prod_{n} 
\left(\sqrt{\frac{m }{2\pi \,\varepsilon}} \,d{x}_n\right)\, \exp\left[
-\frac{m }{2\,\varepsilon} \sum_{n=0}^{N-1}(x_{n+1}-x_n)^2 \right]
\end{align}
in which $L$ is the extension of the 1d box, and $\varepsilon = \beta /N$ with 
$\beta=1/T$ as the inverse temperature  \cite{wipf}. 
The above expression is to be supplemented by the periodic condition $x_0=x_N$ (in the 
continuous-time form $x(0)=x(\beta)$). In the limit $L\to\infty$ Eq.~(\ref{01}) is reduced 
to the well-known expression 
\begin{align}\label{02}
Z_\mathrm{particle}^\mathrm{1d} =(2\pi m/\beta)^{1/2} L
\end{align}
The crucial observation is about the fined-tuned weight-factor
$\displaystyle{\sqrt{\frac{m }{2\pi \,\varepsilon}}}$, which is essential both from 
the dimensional considerations, as well as the expected dependence of the thermodynamical 
quantities on the particle mass. The weight-factor is essential also by expectations from 
the energy-mass relation. The transfer-matrix element between positions at two adjacent 
times $n+1$ and $n$ is defined by 
\begin{align}\label{03}
\langle x'|e^{-\varepsilon \hat{H}} | x \rangle = \langle x',n+1 | x,n \rangle=
\sqrt{\frac{m }{2\pi \,\varepsilon}}  \exp\left[-\frac{m }{2\,\varepsilon}(x'-x)^2 \right]
\end{align}
by which, using the plane-wave $\langle x | p \rangle = \exp(\mathrm{i}\,p\,x)/\sqrt{2\pi}$, 
one finds
\begin{align}\label{04}
\langle x'|e^{-\varepsilon \hat{H}} | x \rangle = \int \frac{dp}{2\pi}\, e^{\mathrm{i}\, p(x'-x)}e^{-\varepsilon E_p}
\end{align}
with 
\begin{align}\label{05}
E_p=\frac{p^2}{2m}
\end{align}
%We see that the specific appearance of the mass $m$ in the weight-factor leads to the expected 
%dependence of the energy on the mass. In particular, and in quite relation with the lattice model we will consider, 
%at fixed momentum and in the limit $m\to \infty$ we have $E_p\to 0$. 
This little exercise shows that the fixing of the normalization pre-factor in the definition of the 
transfer-matrix is quite essential to meet the expectations by the relevant physics, especially the 
dependence of the spectrum on the defining parameters of the model, in this case mass. 

Now let us come back to the lattice gauge model, for which we consider the simplest model of 2d pure 
U(1) model, defined in the temporal gauge $A^0\equiv 0$ by the Euclidean action \cite{wilson,wipf}
\begin{align}\label{06}
S_E= -\frac{1}{g^2} \sum_n \sum_k
\Big(1-\cos\big[a\,g\big(A_{n+1,k}-A_{n,k}\big)\big]\Big),~~~~~~
-\frac{\pi}{a\,g}\leq A \leq \frac{\pi}{a\,g}
\end{align}
in which $n$ and $k$ are labeling lattice links in time and space directions, respectively. 
In the above `$\,a\,$' is the lattice spacing parameter, and
$g$ is the dimensionless gauge-coupling.
The partition function per-link of 2d U(1) model 
can be evaluated by \cite{wilson, wipf}
\begin{align}\label{07}
Z_\textrm{link}(g)=
\int_{-\pi/(a\,g)}^{\pi/(a\,g)}\, \prod_{n} \frac{a\,d A_{n}}{\sqrt{2\pi}}
\,\exp\left[ -\frac{1}{g^2} \sum_{n}
\Big(1-\cos\big[a\,g\big(A_{n+1}-A_{n}\big)\big]\Big)\right]
\end{align}
in which `$\,a\,$' in the measure is inserted to make the above dimensionless.
However, as far as the dimension of the above expression is concerned, any function
of the dimensionless gauge-coupling $g$ can be inserted in the measure too. 
As seen in the example with particle dynamics, the normalization pre-factor
directly affects the dependence of the spectrum on the defining parameters of the model, 
in this case, the gauge-coupling $g$. We shortly come back to this issue and explore 
to some extent the implications of different appearances of $g$ on the spectrum of the model. 
For now, this explanation might be satisfactory that, as in the continuum limit $agA\ll 1$, 
using $\cos\delta\approx 1-\delta^2/2$, there is no $g$ in the exponent, in comparison 
with the particle dynamics example, we expect no $g$ in the measure. 

In the lattice gauge model, it is quite common to change the gauge variables to the angle ones, 
defined by $\theta=a\,g A$ with $-\pi\leq \theta \leq \pi$ \cite{wilson}. The change in the integration variable then 
leads to 
\begin{align}\label{08}
Z_\textrm{link}(g)=
\int_{-\pi}^{\pi}\, \prod_{n} \frac{d \theta_{n}}{\sqrt{2\pi}\,g}
\,\exp\left[ -\frac{1}{g^2} \sum_{n}
\Big(1-\cos\big(\theta_{n+1}-\theta_{n}\big)\Big)\right]
\end{align}
in which the appearance of $1/g$ in the measure is remarked. 
By the small-$\theta$ expansion, the appearance of $1/g$ is reasonable as 
it appears as the mass in (\ref{03}).
By above, the corresponding transfer-matrix 
$\hat{V}=\exp(-a\hat{H})$ of the model is defined by its element
between the angle variables in two adjacent times 
\begin{align}\label{09}
\langle \theta' |\hat{V}_\mathrm{link} | \theta\rangle =
\langle \theta',n+1 | \theta,n\rangle_\mathrm{link}=
 \frac{1}{\sqrt{2\pi}\,g}
  \exp\!\left[\frac{-1}{g^2}\Big[1-\cos\big(\theta'-\theta\big)\Big]\right]
\end{align}
The above has an extra $1/g$ normalization pre-factor in comparison to the 
\textit{commonly} used expression for the matrix element \cite{wipf}. It is in this respect that the commonly
used definition of the transfer-matrix for the lattice gauge models is interpreted at the beginning of this work as 
the one with ``neglected normalization''. It is crucial to see the consequences of this extra $1/g$ factor on the 
spectrum and in general, the one with arbitrary power $\alpha$ 
\begin{align}\label{10}
\langle \theta' |\hat{V}_\mathrm{link} | \theta\rangle_\alpha =
 \frac{1}{\sqrt{2\pi}\,g^\alpha}
  \exp\!\left[\frac{-1}{g^2}\Big[1-\cos\big(\theta'-\theta\big)\Big]\right]
\end{align}
The exact spectrum of the 2d pure U(1) lattice gauge model is quite known 
by means of the transfer-matrix method \cite{wipf}, by which
the eigenvalues of the Hamiltonian and the transfer-matrix are 
related $E=-\frac{1}{a}\ln v$. 
It is easy to see that in the present model the transfer-matrix and the Hamiltonian are diagonal in 
the plane-wave Fourier basis \cite{wipf,spchfath,mattis}
$\langle\theta|s \rangle=\frac{1}{\sqrt{2\,\pi}}
\,\exp(\ri\,s\,\theta)$, for integer $s$.
Using the identity for $\I_s(x)$ as the modified Bessel function of the first kind
\begin{align}\label{11}
\exp(x\,\cos\theta)=\sum_s\I_s(x)\,\exp(\ri\,s\,\theta)
\end{align}
and the relation 
$\int_{-\pi}^\pi\rd\theta\,\exp(\ri\,r\,\theta)=2\pi\,\delta_{0,r}$,
one directly finds the matrix elements of $\hat{V}_\mathrm{link}$ in the 
Fourier basis \cite{spchfath,mattis}
\begin{align}\label{12}
\langle{s}'|\hat{V}_\mathrm{link}|{s}\rangle_\alpha=\frac{\sqrt{2\pi}}{g^\alpha}\,
e^{-1/g^2}\, \,
 \I_{s}\!\left(\frac{1}{g^2}\right)\, \delta_{s's}
\end{align}
by which the energy per-link is found:
\begin{align}\label{13}
E_s= -\frac{1}{a}\left[\ln \left(\frac{\sqrt{2\pi}}{g^\alpha}\right) -\frac{1}{g^2}
+ \ln \I_{s}\!\left(\frac{1}{g^2}\right)\right]
\end{align}
By the property $\I_0> \I_{s\neq 0}$, the ground-state is by ${s}={0}$ \cite{wipf}. 
Using the asymptotic behavior for large arguments of Bessel functions 
in the saddle-point approximation
\begin{align}\label{14}
\I_s(\gamma)\simeq \frac{e^\gamma}{\sqrt{2\pi \gamma}}\,
e^{-(s^2-1/4)/(2\gamma)}\,\left[1+\mathrm{O}\left(\frac{1}{\gamma^2}\right)\right]
,~~~~~~~~~~~\frac{1}{g^2}=\gamma\gg s
\end{align}
one may easily find the spectrum at small coupling limit:
\begin{align}\label{15}
E_s\simeq \frac{\alpha-1}{a}\ln g+ \frac{g^2}{2\,a}\left(s^2-\frac{1}{4}\right),~~~~~~~ g\ll 1
\end{align}
by which the following behaviors are expected
\begin{equation}\label{16}
\alpha>1:\left\{\begin{aligned}
&E_s\to -\infty, ~~ \mathrm{for}~g\to 0 \\
&\frac{\partial E_s}{\partial g} >0,~~~~ g\ll 1, ~\text{all}~s
\end{aligned} \right.
\end{equation}
and
\begin{equation}\label{17}
\alpha<1:\left\{\begin{aligned}
&E_s\to +\infty, ~~ \mathrm{for}~g\to 0  \\
&\frac{\partial E_s}{\partial g} <0,~~~~ g\ll 1, ~\text{all}~s
\end{aligned} \right.
\end{equation}
\begin{figure}[t]%[!ht]
	\begin{center}
		\includegraphics[scale=.65]{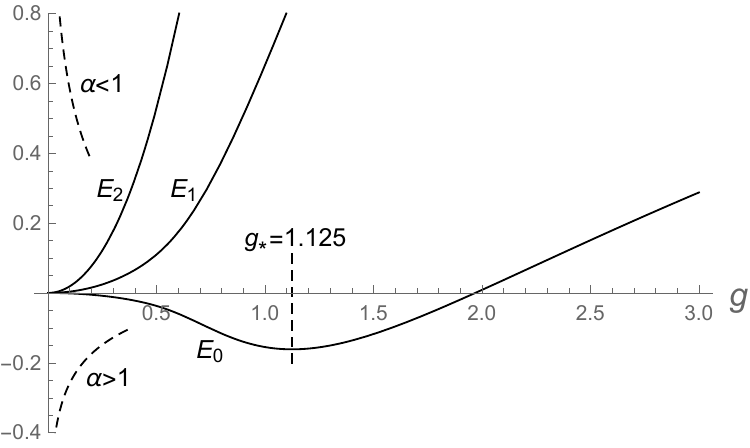}
	\end{center}
	\caption{\small The plots of the three lowest energies by the 
$g$-dependent $\alpha=1$ of (\ref{18}). 
The dashed curves represent the typical behavior of energies at $g\ll 1$ for $\alpha<1$ and $\alpha >1$.}
\end{figure}
By the choice (\ref{16}), we see that the spectrum is in fact 
unbounded from below in extreme weak coupling limit.
The commonly used choice of constant normalization ($g$-independent 
with $\alpha=0$) is in fact a special case of (\ref{17}) \cite{wipf}.
Now by the choice $\alpha=1$, the lowest energy $E_0$ develops a minimum, 
leading to the behaviors 
\begin{equation}\label{18}
\alpha=1: ~\left\{\begin{aligned}
& E_s \to 0,~~~~~~~~~\mathrm{all}~s, ~g\to 0  \\
&\frac{\partial E_0}{\partial g} < 0, ~~~~~~~~   g< g_* \\
&\frac{\partial E_0}{\partial g} > 0, ~~~~~~~~   g> g_*   \\
&\frac{\partial E_s}{\partial g} > 0,~~~~~~~~    s\neq 0,~\mathrm{all}~g
\end{aligned}
 \right.
\end{equation}
with  $g_*=1.125$. The plots of the few lowest energies by the choice 
(\ref{18}) are presented in Fig.~1. 
The appearance of the above minimum is found in \cite{spchfath}, in which 
a spin-chain interpretation of the present model for worldline 
of a particle with the effective-mass $1/(a\,g^2)$ 
is considered. In \cite{spchfath,pvfath} it is discussed 
how the minimum leads to a first-order phase transition by the model. 
As it will be reviewed shortly in here, the very same phase transition observed in 
\cite{spchfath,pvfath} is exhibited by the present lattice gauge model 
with the $\alpha=1$ choice as well. 
As the present 2d pure U(1) lattice gauge model is known
as a single-phase model \cite{wipf}, 
the mentioned phase transition is interpreted as the ``lost in normalization" by the present work's title. 
In fact, the conclusion that the present model is a single phase one is originated by the commonly used 
$g$-independent normalization $\alpha=0$. 
‎\begin{figure}[t]‎
‎	\begin{center}‎
	‎	\includegraphics[scale=0.7]{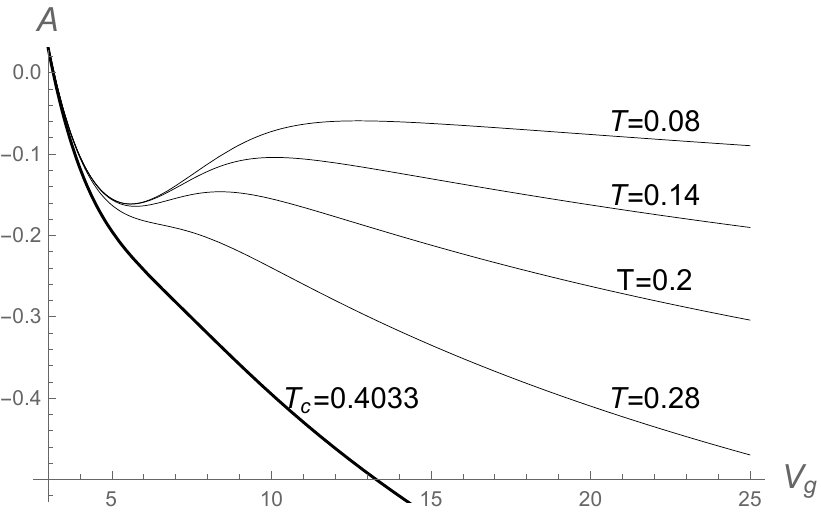}‎
‎	\end{center}‎
‎%\vskip‎ -‎.5cm‎
‎	\caption{\small The isothermal $AV_g$-diagrams at 
different temperatures for the 2d U(1) model‎.}
‎\end{figure}‎

As mentioned earlier, the phase transition is the consequence of the appearance of the 
minimum in the ground-state. At sufficiently low temperatures, the ground-state finds the major 
contribution to the partition function, by which it is expected that the 
isothermal free-energy $A=-T\ln Z$ curves have 
points with common tangents (slopes) below some critical temperature $T_c$. 
The situation is exactly similar to the gas-liquid system, in which 
due to the minimum in the inter-atomic potential, the isothermal $AV$-plots
would find points with common slopes ($V$ as volume) \cite{Huang}. 
In the present case the extent (volume) in the field-space is
$V_g=2\pi/(ag)$ by (\ref{06}). In Fig.~2, we see clearly the same repeated slopes
in the isothermal $AV_g$-curves of the U(1) model with normalization 
(\ref{18}) for temperatures below $T_c=0.4033$. 
In the gas-liquid system, the conjugate variable of volume $V$ is the pressure 
defined by $P=(\partial A‎/ ‎\partial V)_T$, by which the Gibbs energy of 
the gas-liquid system reads $G=A+P\,V$. 
In the present case the thermodynamical conjugate variable of coupling $V_g$ 
as $P_g$ is similarly defined as $P_g=-(\partial A‎/ ‎\partial V_g)_T$‎, 
by which the Gibbs energy is defined as $G=A+P_g\, V_g$.
Similar to the case of the gas-liquid system for $GP$-diagrams 
\cite{Huang,stanley}, the mentioned behavior of the free-energy‎ would
cause that ‎the Gibbs energy would be multi-valued as a function of $P_g$, 
leading to cusps in the isothermal $GP_g$-diagrams below $T_c$‎. 
‎For the present U(1) model,‎ the ‎plots of the isothermal $GP_g$-curves
are presented in Fig.~3‎, ‎in which the expected cusps are evident‎.
In the case of the gas-liquid system, it is known that at equal
$P$ and $T$ ‎the state with lower $G$‎ ‎is selected by the system
leading to the jump in the derivative $\partial G/\partial P$ at cusps
\cite{Huang,stanley}. As a result, the pressure remains constant in 
the system during a first-order phase transition.
In the present 2d U(1) model, the same lower-$G$ rule saves the 
system from having multi-valued $G$, at the cost of undergoing 
a first-order phase transition due to the jump in the derivative of $G$.

‎\begin{figure}[t]‎
‎	\begin{center}‎
	‎	\includegraphics[scale=0.55]{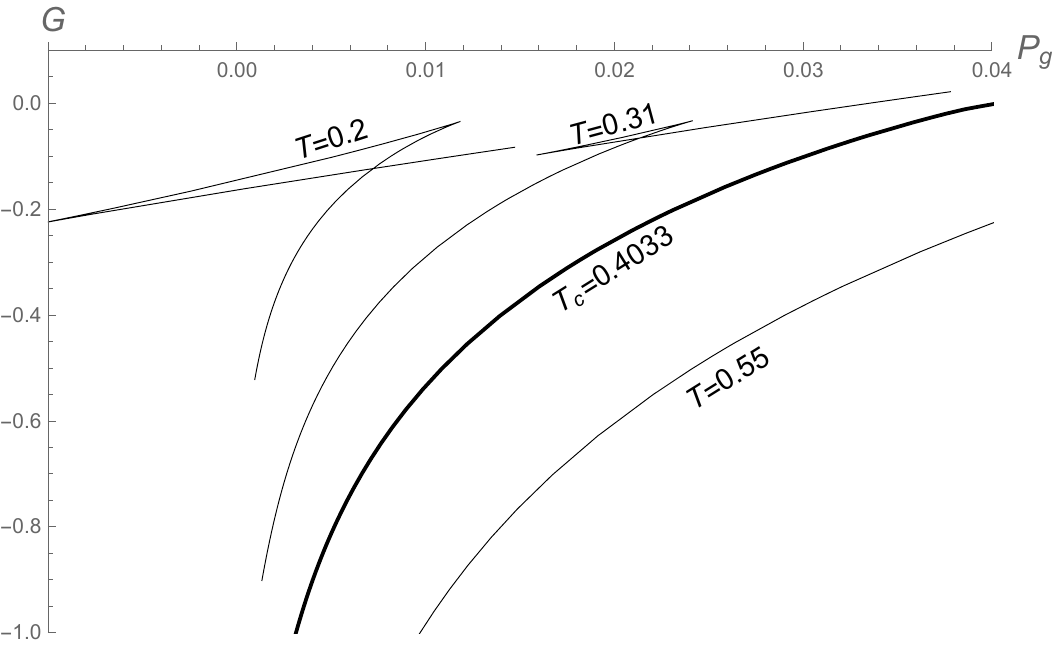}‎
‎	\end{center}‎
%‎\vskip‎ -‎.5cm‎
‎	\caption{\small The isothermal $GP_g$-diagrams for the 2d U(1) model‎ by $\alpha=1$ choice.}‎
‎\end{figure}‎

As is discussed in detail in \cite{pvfath}, the mentioned phase transition 
is a direct consequence of the appearance of the model's parameter in the normalization,
leading to the minimum in ground-state by the $\alpha=1$ choice of (\ref{18}). 
To make a judgment between different $g$-dependent normalizations,
the energy spectrum of the model at the continuum limit is considered.
Quite remarkably, based on a semi-classical treatment of the model, it is found that at the next-leading order
of the continuum limit the spectrum is compatible only with the $\alpha=1$ choice, especially the 
negative slope of $E_0$ at small $g$. 

Our standing point to make a judgment between different 
normalizations is the continuum limit $a\to 0$, 
leading to the following replacements 
\begin{align}\label{19}
\Delta \theta &= \theta_{n+1}- \theta_n \to a\,\dot{\theta} +\frac{1}{2!}a^2\,\ddot{\theta} 
+\frac{1}{3!}a^3\,\dddot{\theta} +\cdots\\
\label{20}
\cos\Delta \theta&\simeq 1- \frac{1}{2!}\,\Delta \theta^2
+\frac{1}{4!}\,\Delta \theta^4+\cdots\\
\label{21}
\sum_{n} &\to a^{-1}\int dt
\end{align}
Using $\,\dot{\theta}\,\dddot{\theta}=d(\dot{\theta}\ddot{\theta})/dt- \ddot{\theta}^2\,$ and 
after dropping the surface terms, an action consisting the 2nd-derivative is obtained. Back from 
imaginary-time to the real one, the action per-link takes the form 
\begin{align}\label{22}
S=\frac{a}{g^2} \int \!\! dt ~\left[\frac{1}{2}\, \dot{\theta}^2+\frac{1}{24} a^2
(\ddot{\theta}^2 +\dot{\theta}^4)+\mathrm{O}(a^4)
\right]
\end{align}
First, let us find the spectrum at the leading order, for which 
only the first term as the kinetic term is concerned. 
The canonical momentum at this order is given by 
\begin{align}\label{23}
p:=\frac{\partial L}{\partial \dot{\theta}} =\frac{a}{g^2}\, \dot{\theta} + \cdots
\end{align}
leading to the Hamiltonian
\begin{align}\label{24}
H=\frac{g^2}{2a}\,p^2+ \cdots
\end{align}
By the fact $-\pi\leq \theta \leq \pi$, in the quantum theory the momentum
takes the discrete values $p=s$ with $s\in\mathbb{Z}$,
leading to the energy spectrum per-link
\begin{align}\label{25}
E_s =\frac{g^2}{2\,a}\,s^2+ \cdots
\end{align}
The above, apart from the ``$-1/4$" term that we will come back to it very soon, agrees with 
the expression (\ref{15}) with $\alpha=1$. The following is needed to be held
\begin{align}\label{26}
\frac{g}{a}=\mathrm{finite~~as~~~} a\to 0
\end{align}
by which $g^2/a\to 0$, to gain a finite continuous spectrum in the 
limit $a\to 0$. In fact, the expectation that the continuum limit of lattice theory 
reflects the small coupling limit \cite{wilson} is translated by (\ref{26}).
Also, the approximations (\ref{19}) and (\ref{20}) suggest the following 
agreement of orders
\begin{align}\label{27}
a\,\dot{\theta} \sim \eta \, g \ll 1
\end{align}
in which $\eta$ is a dimensionless number of order one. 
Now let us go beyond the leading order in the continuum limit. 
The action (\ref{22}) consists of the 2nd order time-derivative, for which it is known that the 
Hamiltonian formulation is due to Ostrogradsky \cite{ost1,ost2,ost3}. Accordingly, the
phase space variables are defined as \cite{ost1,ost2,ost3}:
\begin{align}\label{28}
q:=&\theta,~~~~~~p:=\frac{\partial L}{\partial\dot{\theta}}
-\frac{d}{dt}\frac{\partial L}{\partial\ddot{\theta}},\\
\label{29}
q':=&\dot{\theta},~~~~~~p':=\frac{\partial L}{\partial\ddot{\theta}}
\end{align}
by which the following canonical relations hold \cite{ost1,ost2,ost3}
\begin{align}\label{30}
\{q,p\}=\{q',p'\}=1, 
\end{align}
with others being zero. Provided the Hamiltonian is defined as follows \cite{ost1,ost2,ost3}
\begin{align}\label{31}
H=\dot{q}\,p+\dot{q'}\,p' - L
\end{align} 
the 4th-order equation of motion is recovered
\begin{align}\label{32}
\frac{d^2}{dt^2}\frac{\partial L}{\partial\ddot{\theta}}-
\frac{d}{dt}\frac{\partial L}{\partial\dot{\theta}}
+\frac{\partial L}{\partial{\theta}}=0
\end{align}
in which the last term vanishes as the Lagrangian does not depend 
explicitly on $\theta$, leading to conserved $p$ of (\ref{28}). Also,
as the Lagrangian does not explicitly depend on time, the energy $E$ 
is represented by the Hamiltonian and is conserved \cite{ost1,ost2,ost3}. 
For the present case, we find explicitly
\begin{align}\label{33}
E&=\frac{a}{g^2}\left(  \frac{1}{2}\dot{\theta}^2 +\frac{1}{8}a^2 \dot{\theta}^4+
\frac{1}{24}a^2\ddot{\theta}^2 - \frac{1}{12}a^2 \dot{\theta}\,\dddot{\theta}
+\mathrm{O}(a^3)
\right)\\
\label{34}
p&=\frac{a}{g^2}\left(\dot{\theta} +\frac{1}{6}a^2\dot{\theta}^3
-\frac{1}{12}a^2 \dddot{\theta}+\mathrm{O}(a^3)\right).
\end{align}
As both $p$ and $E$ are conserved, they are determined by the initial conditions:
\begin{align}\label{35}
\dot{\theta}(0),~~~\ddot{\theta}(0),~~~\dddot{\theta}(0)
\end{align}
By the square of $p$, one can replace the combination of 
$ \dot{\theta}\,\dddot{\theta}$ in the energy expression 
by which, after setting $p=s$ with $s\in\mathbb{Z}$ as
the conjugate momentum of the compact variable 
$-\pi\leq\theta\leq\pi$, one find for the energy 
\begin{align}\label{36}
E_s=\frac{g^2}{2\,a}\,s^2 + \frac{a^3}{24\,g^2}(\ddot{\theta}^2-\dot{\theta}^4)+\mathrm{O}(a^4)
\end{align}
in which the first term matches to the leading order result (\ref{25}). 
By (\ref{36}), the lower energy is obtained by the initial condition $\ddot{\theta}(0)= 0$,
by which the positive term $\ddot{\theta}^2$ is absent.  
The domain of validity (\ref{27}) can be used to obtain the lowest possible
energy by fixed $s$ at this order, namely 
\begin{align}\label{37}
E_s^\mathrm{min} = \frac{g^2}{2\,a}\left(s^2 - \frac{\eta^4}{12}\right),~~~g\ll 1
\end{align}
which matches with the expression (\ref{15}) with $\alpha=1$, suggesting the 
value $\eta=3^{1/4} \simeq 1.32$, being order of unity as expected. 
The remarkable fact by (\ref{37}) is about the slopes at the limit $g\ll 1$
\begin{align}\label{38}
&\frac{\partial E_0^\mathrm{min}}{\partial g^2} = -\frac{\eta^4}{24\,a} <0\\
\label{39}
&\frac{\partial E_{s\neq 0}^\mathrm{min}}{\partial g^2} = 
\frac{1}{2\,a} \left(s^2-\frac{\eta^4}{12}\right) >0
\end{align}
We saw earlier that the behaviors (\ref{38}) and (\ref{39}) are 
recovered only by the choice (\ref{18}) for the normalization factor. 

In conclusion, the consequences of the coupling-dependent normalizations 
in the definition of the transfer-matrix of lattice gauge theories are explored.
As an illustrative example, the exactly solvable 
model of 2d pure U(1) lattice gauge theory is considered to 
explore the mathematical and physical consequences of 
different normalizations of transfer-matrix. Among the power-law normalizations $1/g^\alpha$, 
it is observed that the power $\alpha=1$ matches 
the semi-classical spectrum by the model. In particular, it is found that at the 
small coupling limit $g\ll 1$, the lowest energy has the decreasing behavior 
$E_0\!\propto\! -g^2$ while the higher energies are increasing as 
$E_{s\neq 0}\!\propto\! g^2$. It is seen that in fact, this is the 
case by normalization (\ref{18}) for 2d U(1) model, by which a minimum 
in ground-state is developed at $g_*=1.125$.
The thermodynamical consequences of the mentioned minimum 
are reviewed \cite{spchfath,pvfath}. In particular, it is discussed how due to the minimum 
in the ground-state the Gibbs energy $G$ would be a multi-valued function in terms of 
the conjugate variable of $V_g=2\pi/g$ at constant $T$. The similar 
behavior for the gas-liquid system is mentioned, leading to a first-order transition
between two phases \cite{Huang,stanley}. The exact nature of the phase transition in the present 
2d U(1) model remains to be understood. 

As the final remark, it is expected that the similar decreasing behavior of 
the ground-state energy at small $g$ would be observed
for lattice gauge theories in higher dimensions as well. 
This simply comes back to the fact that by the higher-order
time-derivatives, due to the Ostrogradsky construction, an
opposite slope for the lowest possible energy at small $g$ is expected.
However, it is expected that a careful tuning of the power of $g$ in 
normalization factor is needed. This is because that in higher dimensions, 
besides the spatial $A_n$ of the present model, there are still redundant degrees 
of freedom, as the gauge is not totally fixed by the temporal gauge used in here. 

\vskip .5cm
\textbf{Acknowledgment}
The authors are grateful to M. Khorrami for useful discussions. 
This work is supported by the Research Council of Alzahra University.


\begin{thebibliography}{99}

\bibitem{wipf} A. Wipf, ``Statistical Approach to Quantum Field Theory", 
Springer 2013, Chaps. 13 \& 14.

\bibitem{wilson} K.G. Wilson, 
``Confinement of Quarks",
Phys. Rev. D \textbf{10} (1974) 2445.

\bibitem{mattis} D.C. Mattis, ``Transfer Matrix in Plane-Rotator Model",
Phys. Lett. A \textbf{104} (1984) 357.

\bibitem{spchfath} A.H. Fatollahi, 
``Worldline as a Spin Chain", 
Eur. Phys. J. C \textbf{77} (2017) 159,
1611.08009 [hep-th].

\bibitem{pvfath} A.H. Fatollahi, ``First-Order Phase Transition by the XY Model of Particle Dynamics", 
Europhys. Lett. \textbf{128} (2019) 27002, 1811.02408 [stat-mech]

\bibitem{Huang} K‎. ‎Huang‎, ‎``Statistical Mechanics"‎, ‎Wiley 1987‎. 

\bibitem{stanley} H.E‎. ‎Stanley‎, ‎``Introduction to Phase Transitions and Critical Phenomena"‎, 
‎Oxford Univ‎. ‎Press 1971‎, ‎Sec.~2.5‎. 

\bibitem{ost1} M. de Leon, P.R. Rodrigues
``Generalized Classical Mechanics and Field Theory: A Geometrical Approach 
of Lagrangian and Hamiltonian Formalisms Involving Higher Order Derivatives",
Elsevier 2011.

\bibitem{ost2} R.P. Woodard, Lecture Notes in Physics. \textbf{720} (2007) 403–433, 
astro-ph/0601672;\\
``The Theorem of Ostrogradsky", 1506.02210[hep-th].

\bibitem{ost3} J. Govaerts and M.S. Rashid, ``The Hamiltonian Formulation of Higher Order Dynamical Systems", hep-th/9403009.

\end{thebibliography}
\end{document}